\documentclass[epj]{webofc}
\usepackage[varg]{txfonts}   
\usepackage{graphicx}  
\usepackage{dcolumn}   
\usepackage{bm}        
\usepackage{amssymb}   
\usepackage{amsmath,amsfonts} 
\usepackage{slashed}  
\usepackage{rotating}
\usepackage{overpic}
\usepackage[absolute,overlay]{textpos}
\usepackage{xcolor}
\usepackage{epstopdf,psfrag}
\wocname{EPJ Web of Conferences}
\woctitle{CONF12}
\newcommand{\unit}{1\!\!1}
\DeclareMathOperator{\e}{e}
\DeclareMathOperator{\img}{i}

\DeclareMathOperator{\su}{SU}
\DeclareMathOperator{\dd}{d}

\begin{document}
\selectlanguage{english}
\title{Quark Propagator with electroweak interactions in the Dyson-Schwinger approach}
%
%

\author{Walid Ahmed Mian\inst{1}\fnsep\thanks{\email{walid.mian@uni-graz.at}} \and
        Axel Maas\inst{1}\fnsep\thanks{\email{axel.maas@uni-graz.at}}
}

\institute{Institute of Physics, NAWI Graz, University of Graz,
Universit\"atsplatz 5, A-8010 Graz, Austria}

\abstract{%
  Motivated by the non-negligible dynamical backcoupling of the electroweak interactions with the 
  strong interaction during neutron star mergers, we study the effects of the explicit breaking of 
  C, P and flavor symmetry on the strong sector. 
  The quark propagator is the simplest object which encodes the consequences of these breakings.
  
  To asses the impact, we study the influence of especially parity violation on the propagator for various masses. 
  For this purpose the functional methods in form of Dyson-Schwinger-Equations are employed. 
  We find that explicit isospin breaking leads to a qualitative change of behavior even for 
  a slight explicit breaking,  which is in contrast to the expectations from perturbation theory. 
  Our results thus suggest that non-perturbative backcoupling effects could be larger than expected.
}
\maketitle
%

\section{Introduction}

One of the most remarkable observations of this century so far is the one of gravitational waves \cite{Abbott:2016blz}.
This has opened a new window to investigate the universe. One target for investigations using gravitational waves are binary neutron star mergers.
Various investigations show that the form of the gravitational waves is influenced by the electroweak interactions,
due to the dynamical back coupling with the strong interaction in such dense environments, see 
\cite{Rosswog:2003rv,Sekiguchi:2011zd,Neilsen:2014hha,Palenzuela:2015dqa,Sekiguchi:2015dma,Foucart:2015gaa,Caballero:2016lof} 
and references therein.

A first-principles study of this requires a fully coupled, non-perturbative treatment of this system. 
However, due to the enormous complexity a fully backcoupled analysis is too complicated as a first step. 
Thus we focus here on a few particular aspects of the backcoupling of the electroweak interactions to the strong sector. 
These are the explicit breaking of the C, P and flavor symmetry.

Symmetries reduce, by virtue of the Wigner-Eckhart theorem, the structure of the correlation functions. 
Therefore the explicit symmetry breaking induces a richer tensor structure. 
The quark propagator (QP) is the simplest object to exhibit the full additional complexity, 
which we will thus analyze here and discuss its most pertinent features.
A full investigation of this will be available elsewhere \cite{Maas2016prep}.

To do so, we have to cope with very large differences in energy scales and parity violation. 
This limits our choice of possible methods, and makes especially lattice gauge theory not feasible at the moment. 
Rather, we turn to functional methods, especially Dyson-Schwinger equations (DSEs). 
These are a continuous, covariant and non-perturbative formulation and 
allow us to access high and low energies at the same time. 
Their major drawback is that they lead to an infinite tower of coupled non-linear integral equations, 
which need truncations to be solved. Nevertheless these methods yield good results at various 
levels of sophistication to determine the QP, see e.\ g.\ 
\cite{Alkofer:2000wg,Fischer:2006ub,Alkofer:2008tt,Roberts:2015lja,Williams:2015cvx} and references therein.

In this study we restrict ourselves to the so-called rainbow-ladder truncation. 
This truncation conserves the most important features for our purpose, like dynamical mass generation 
and the correct treatment of chiral symmetry \cite{Alkofer:2000wg,Fischer:2006ub,Roberts:2015lja}. 
Further details about our ansatz and the corresponding DSEs are given in section \ref{Basics}.

We find that the mass splitting between the quarks in an isospin doublet leads to a threshold symmetry breaking strength, 
which is still very small, but were we already see a qualitative change in behavior for parity violation. 
Based on perturbation theory, this is not expected. This result will be discussed in section \ref{Parity Violation}.

In the end we summarize our results in section  \ref{Conclusion}.

\section{Basics and Methods}
\label{Basics}

To investigate the dynamical backcoupling from the breaking of C, P and flavor symmetries and avoiding all
the complexities coming from the full electroweak interactions, we insert an explicit breaking 
term in our Lagrangian to keep it as simple as possible. 
Inspired by the propagation of the quarks from one flavor into another in the presence of the electroweak interactions, 
we only consider the flavor breaking for the QPs of the same generation. 
This together with the C and P violation leads to
\begin{align}
  \mathcal{L} =&  \mathcal{L}_{\text{QCD}} +  \mathcal{L}_{\text{Effective}}, 
  \nonumber \\
   \mathcal{L}_{\text{QCD}} =&\overline{\psi}_{u} \left[ - \slashed{\partial} +m_u \right] \psi_u +
   \overline{\psi}_{d} \left[ - \slashed{\partial} +m_d \right] \psi_d +  
   g_s \overline{\psi}_{u} \slashed{A^{i}} T^{i} \psi_u 
   + g_s \overline{\psi}_{d} \slashed{A^{i}} T^{i} \psi_d   +  \mathcal{L}_{\text{Rest}},
  \nonumber \\ 
  \mathcal{L}_{\text{Effective}}= &
   - 2 g_{\text{w}} \left( \overline{\psi}^{L}_{u} \slashed{\partial} \psi^{L}_{d}   +  
  \overline{\psi}^{L}_{d} \slashed{\partial} \psi^{L}_{u} \right),
  \label{Eqn:Ansatz}
  \\  
  \psi^{L} =& \frac{1}{2}(\unit - \gamma^{5})\psi,
  \nonumber
\end{align}
where $\psi_u$ and $\psi_d$ denotes the fields for the up-like quarks and down-like quarks with 
the corresponding left handed fields $\psi^{L}$. 
$g_s$ is the strong coupling and $g_{\text{w}}$ is the effective symmetry breaking strength. 
$m_{u}$ and $m_{d}$ are the bare masses, which are put to $2.3$ MeV for the up quark, 
$4.8$ MeV for the down quark, $4.18$ GeV for the bottom quark and $160$ GeV for the top quark 
at a renormalization point $\mu^{2}=10^{12}$ GeV. 
$A^{i}$ are the gluon fields and $T^{i}$ are the generators of $\su(N_{c}=3)$. 
The gluon self interaction etc.\ are included in the rest part of the Lagrangian $\mathcal{L}_{\text{Rest}}$. 
For simplicity we suppressed all renormalization constants. We use Landau gauge throughout.

The effective Lagrangian $\mathcal{L}_{\text{Effective}}$ violates parity, 
because it only consists of left-handed fields. By charge conjugation, left-handed fermion 
fields are transformed into right-handed anti-fermion fields. 
Because the effective Lagrangian couples left-handed fermion fields with left-handed anti-fermion fields, 
it violates also charge conjugation. 
Therefore it has the desired properties. But in addition to C, P and flavor violation
it also violates charge conservation, which is not violated in the standard model. 
Rather, in the standard model the decay process of a down-quark into an up-quark at low-energies 
happens by emission of an electron and an anti-electron neutrino. 
Thus, the breaking term implies a coupling with a reservoir of leptons, a constant external field. 
In fact, this is well motivated given that our ultimate goal is the study of binary neutron star mergers, 
which provide such a reservoir 
\cite{Rosswog:2003rv,Sekiguchi:2011zd,Neilsen:2014hha,Palenzuela:2015dqa,Sekiguchi:2015dma,Foucart:2015gaa,Caballero:2016lof}.

As already mentioned symmetry breaking leads to a more complex tensor structures 
for the correlation functions and thus the QPs has in addition to the usual vector 
and scalar channels also non-vanishing axial and pseudo-scalar channels. 
Besides this, we also have non-vanishing QPs for mixed flavors. So the QP from flavor $A$ to flavor $B$ is given by
\begin{align}
  P_{AB} (p^2) &= 
  \tilde{A}_{AB}(p^2) \img \slashed{p} + \tilde{B}_{AB}(p^2) \unit + 
 \tilde{C}_{AB}(p^2) \img \slashed{p} \gamma^5 + \tilde{D}_{AB}(p^2) \gamma^5,
\label{Eqn:QP}
 \end{align}
where $\tilde{A}, \tilde{B}, \tilde{C}$ and $\tilde{D}$ are the corresponding dressing functions.

For a better understanding, let us state the QPs at tree-level, which are given by
\begin{align}
P_{0,uu}(p^2)=& \frac{1}{N(p^{2})} \left[ (m_{d}^{2}+(1-2g_{\text{w}}^{2})p^{2}) \img \slashed{p} 
+ m_{u}(m_{d}^{2}+p^{2}) \unit + 2g_{\text{w}}^{2}p^{2} \img \slashed{p} \gamma^{5} \right],
\nonumber \\
P_{0,dd}(p^2)=& \frac{1}{N(p^{2})} \left[ (m_{u}^{2} + (1-2g_{\text{w}}^{2})p^{2}) \img \slashed{p}
+ m_{d} (m_{u}^{2}+p^{2}) \unit +2g_{\text{w}}^{2}p^{2} \img \slashed{p} \gamma^{5} \right],
\label{Eqn:QPTL} \\
P_{0,ud}(p^2)=& \frac{g_{\text{w}}}{N(p^{2})}\left[ (m_{u}m_{d}-p^{2}) \img \slashed{p} 
- (m_{u}+m_{d})p^{2} \unit -  (m_{u}m_{d}+p^{2}) \img \slashed{p} \gamma^{5}
-(m_{u}-m_{d})p^{2} \gamma^{5}  \right],
\nonumber \\
 P_{0,du}(p^2)=& \frac{g_{\text{w}}}{N(p^{2})}\left[ (m_{u}m_{d}-p^{2}) \img \slashed{p} 
 - (m_{u}+m_{d})p^{2} \unit -  (m_{u}m_{d}+p^{2}) \img \slashed{p} \gamma^{5}
 +(m_{u}-m_{d})p^{2} \gamma^{5}\right],
 \nonumber
\end{align}
with
\begin{align}
N(p^{2})=& 
m_{d}^{2} m_{u}^{2} + (m_{u}^{2} + m_{d}^{2}) p^{2}+ (1-4g_{\text{w}}^{2}) p^{4}.
\end{align}
The tree-level propagator reveals the different contributions to the QPs.
For example the scalar channel at tree-level is proportional to the bare masses and thus depends on explicit chiral symmetry breaking.

As will be seen, the most important contribution in our analysis is the contribution to the pseudo-scalar channel, 
which is proportional to the mass splitting between the up-like and down-like quarks. 
At tree-level this contribution is only present in the pseudo-scalar channel, 
but for the full propagator all channels talk to each other and will become sensitive to the mass splitting. 
This will have a huge impact, as shown in section  \ref{Parity Violation}, 

\begin{figure}
  \centering
  \includegraphics[scale=0.35]{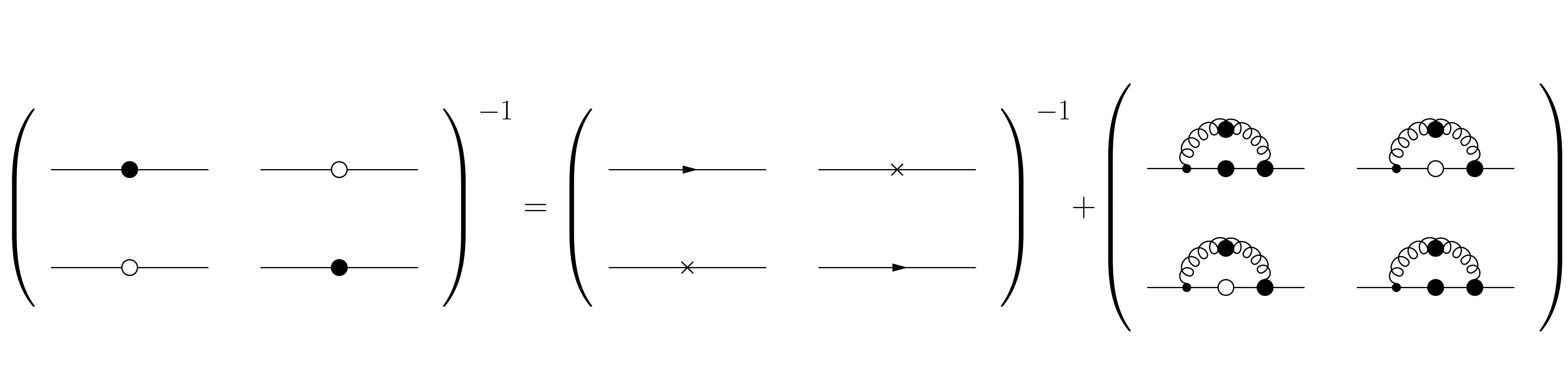}
  \caption{Graphical representation of the DSEs, given as a matrix equation in flavor space. 
  Straight lines are quark propagators, where full circles are full flavor-conserving propagators, 
  open circles are full flavor-mixing propagators, arrows are the tree-level propagators, 
  and crosses are the tree-level breaking terms. 
  Curly lines are gluons, which either couple with a bare vertex (dot) or full vertex (full circle) to the quarks.}
  \label{Pic:DSE}
\end{figure}

The diagrammatic representation of the DSEs is shown in figure \ref{Pic:DSE}. 
On the diagonal, we see the QPs for the pure flavor and on the off-diagonal the QPs for mixed flavors. 
The right-hand side has two contributions. 
The first are the tree-level terms, which include in our ansatz also off-diagonal elements. 
The second terms on the right-hand side come from the QCD interactions. 
If the explicit breaking is switched off, all off-diagonal elements vanish, and ordinary QCD is recovered. 
The full derivation of the DSEs will be discussed in \cite{Maas2016prep}.

To solve these DSEs requires knowledge of the full gluon propagator and the full quark-gluon-vertex, 
which satisfy further DSEs. To deal with this, 
we employ the rainbow truncation \cite{Alkofer:2000wg,Fischer:2006ub,Roberts:2015lja}. 
This truncation replaces the full gluon propagator and the full quark-gluon-vertex 
by the bare quark-gluon-vertex and an effective coupling $\alpha$. 
In this truncation the DSEs reduce to \cite{Maas2016prep}
\begin{align}
 P^{-1}_{AB}(p^2,\mu^2) =& \sqrt{Z_{2,A}(\mu^2,\Lambda^2) Z_{2,B}(\mu^2,\Lambda^2) } P^{-1}_{0,AB} (p^2)
 + \frac{Z_{2,A}(\mu^2,\Lambda^2) Z_{2,B}(\mu^2,\Lambda^2)}{3 \pi^3} \cdot
 \nonumber \\
 & \cdot \int^{\Lambda}{\dd^{4} q \frac{\alpha (k^2)}{k^2}\left(\delta_{\nu \rho} - \frac{k_{\nu}k_{\rho}}{k^2}\right)
 \gamma_{\nu}P_{BA}(q^2,\mu^2) \gamma_{\rho}},
 \label{Eqn:DSE_QP}
\end{align}
where the different $Z$s are the renormalization constants and $k=q-p$. $\int^{\Lambda}$ 
represents a translationally-invariant regularization with the UV-cutoff $\Lambda$. 
It remains to specify $\alpha$.
We choose the Maris-Tandy coupling as an effective coupling, as it has the right 
perturbative behavior at large momenta and ensures dynamical mass generation \cite{Maris:1999nt}. 
It is given by
\begin{align}
   \alpha(q^2) =& \frac{\pi}{\omega^6} D q^4 \e^{-\frac{q^2}{\omega^2}}
   + \frac{2\pi \gamma_m \left[1-\exp{\left(-\frac{q^2}{m_t^2}\right)}\right]}{\ln\left[\e^2-1+\left(1+\frac{q^2}{\Lambda^2_{\text{QCD}}}\right)^2\right]},
  \label{Eqn:MTC}
\end{align}
with the parameters $m_t = 1.0$ GeV, $\omega = 0.4 $ GeV, $D=0.93$ $\mathrm{GeV}^2$ and
$\Lambda_{\text{QCD}} = 0.234$ GeV fixed to the properties of the pion as in \cite{Maris:1999nt,Goecke:2011pe}. 
The anomalous dimension $\gamma_{m}=12/(11 N_{c}-2 N_{f})$ is set for $N_f=2$ and $N_c=3$, 
because each quark generation is considered on its own. 
The numerical solutions of these equations can be performed using standard methods, 
and will not be detailed here further \cite{Maas2016prep}.

\section{Parity Violation}
\label{Parity Violation}

\begin{figure}
 \includegraphics[scale=0.9]{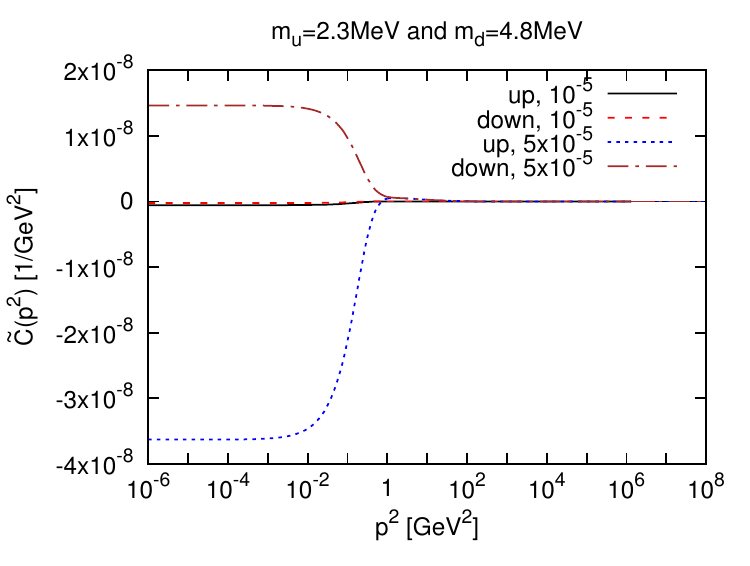}
  \quad
 \includegraphics[scale=0.9]{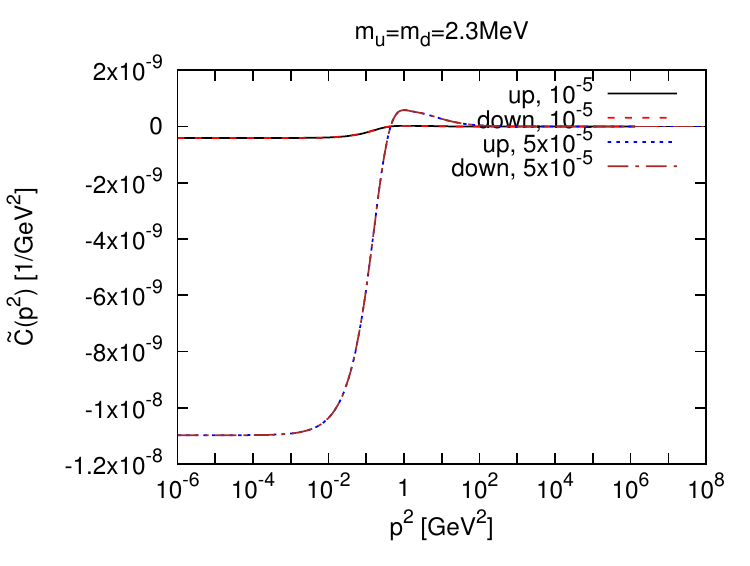}
  \quad
 \includegraphics[scale=0.9]{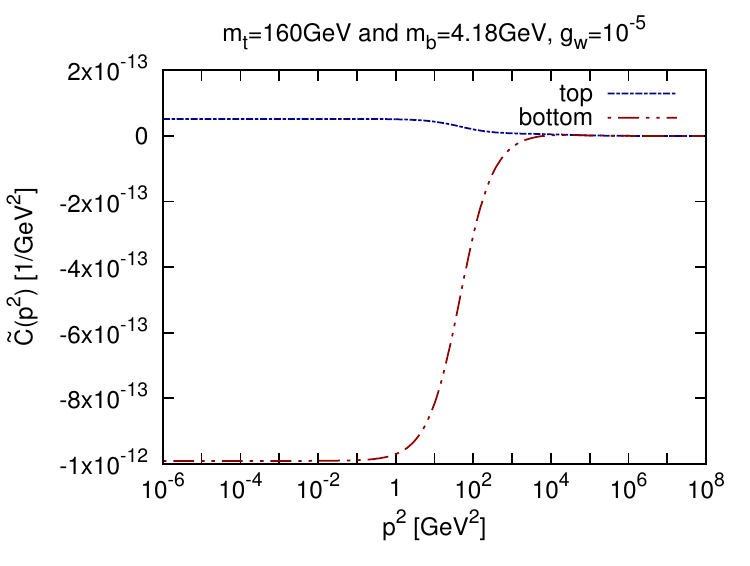}
  \quad
 \includegraphics[scale=0.9]{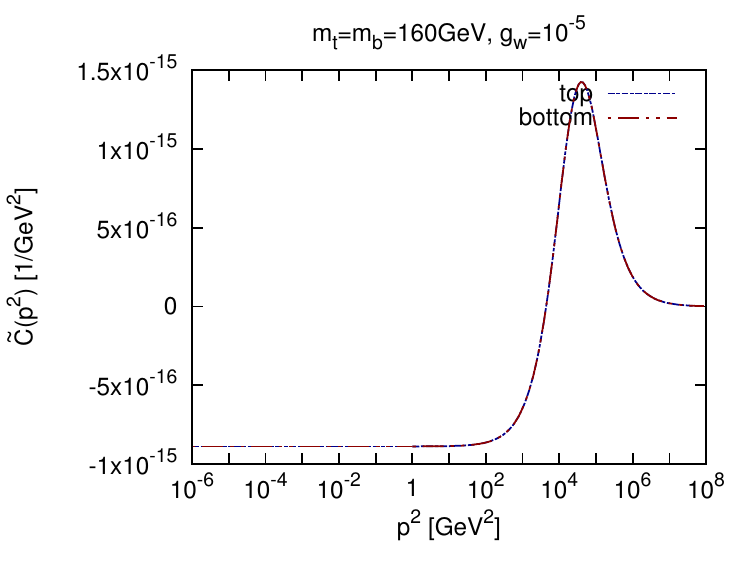}
  \caption{The axial channel for the up and down quark (top panels), and for the top and bottom quark (bottom panels). 
  For the physical quark masses (left-hand side), the heavier quark of the same generation (down and top) have a 
  qualitative  change of behavior in the IR above the threshold strength: 
  $\tilde{C}$ remains positive for all scales in contrast to the other quarks, 
  where a transition scale exists for a sign change. 
  The change of behavior is due to the mass splitting between both quarks, which is shown on the right hand side: 
  If both quarks have the same mass $\tilde{C}$ changes again its sign. 
  The threshold strength is decreased for higher values of the mass splitting, 
  which is seen in $\tilde{C}$ for top and bottom. 
  The change of behavior in this case is also present at $g_{\text{w}}=10^{-5}$, 
  where for up and down no change of behavior is visible.}
  \label{Pic:CP}
\end{figure}

The appearance of the axial- and pseudo-scalar channel is due to parity violation. 
Starting therefore with the axial channel, its tree-level value is, from (\ref{Eqn:QPTL}),
\begin{align}
 \tilde{C}_{0,AA} &= \frac{2g_{\text{w}}^{2}p^{2}}{N(p^{2})}.
 \label{Eqn:TLC}
\end{align}
Hence $\tilde{C}$ has to be positive in the UV because of asymptotic freedom. 
The full result, shown in figure \ref{Pic:CP}, obeys this.

However, something interesting happens at low momenta. 
For sufficiently small breakings, $g_{\text{w}}\lesssim 10^{-5}$, 
there exists a transition scale at around 1 GeV for up and down quarks where $\tilde{C}$ changes sign. 
Starting from a slightly larger coupling $g_{\text{w}}\approx 5\times 10^{-5}$, 
this is no longer true for the down quark, where $\tilde{C}$ remains positive for all momenta. 
Thus there is a threshold strength, where $\tilde{C}$ changes its qualitative behavior for the down quark. 
This change of behavior is given by the mass splitting between the up and down quark. 
This can be seen also in figure \ref{Pic:CP}, since for the mass degenerate case this does not occur. 
There the transition scale for the sign change exists for all investigated strengths.

For the top and bottom quark we already see the change of behavior at $g_{\text{w}}\approx 10^{-5}$, 
and thus a lower value. This can be understood already from the pseudo-scalar channel of the tree-level propagator (\ref{Eqn:QPTL}). 
The contribution, which changes the behavior, is proportional to the mass splitting between the quarks and the strength. 
Thus for top and bottom quark, where the absolute value of the mass splitting is increased, 
less strength is needed for the same effect. 
At the same time the transition scale is shifted towards higher values. 
This is due to higher bare quark masses for top and bottom. 
This effect could be triggered by the vector channel of the mixed QPs. 
They are given at tree-level by
\begin{align}
 \tilde{A}_{0,ud} &= \frac{g_{\text{w}}(m_u m_d - p^{2})}{N(p^{2})}
 \label{Eqn:TLAM}
\end{align}
and thus have contributions in different directions for $p^2 < m_u m_d$ and $p^2 > m_u m_d$. This seems to cause the shifting towards higher values of the transition scale for heavier quarks.

\begin{figure}
 \includegraphics[scale=0.9]{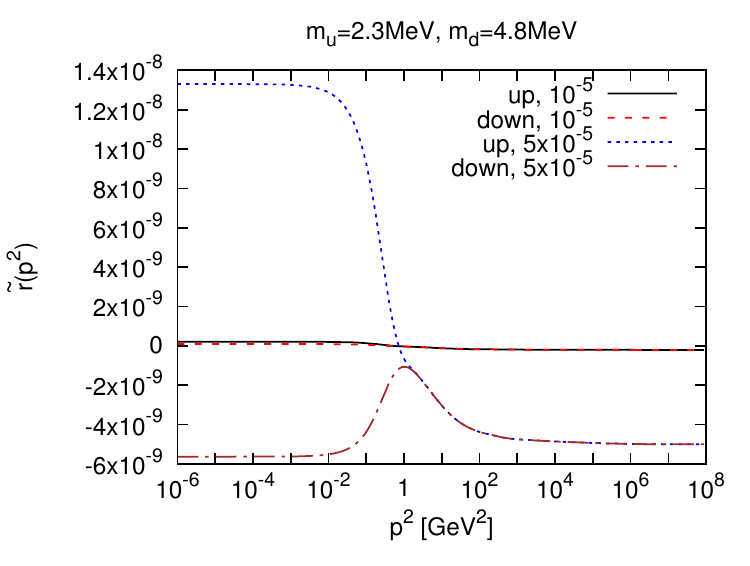}
  \quad
 \includegraphics[scale=0.9]{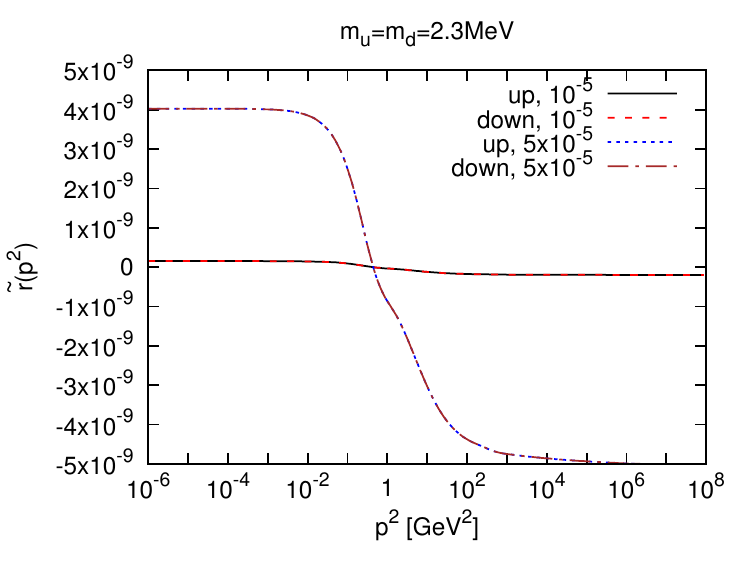}
  \caption{Relative ratio of the left-handed contributions for the physical quark masses (left-hand side) 
  and degenerate masses (right-hand side).  
  For $g_{\text{w}}=10^{-5}$ QPs for up and down have dominantly right-handed contributions 
  in the UV and dominantly left-handed in the IR. 
  For $g_{\text{w}}=5\times 10^{-5}$ the down quark QP changes its behavior by keeping dominantly 
  right-handed contributions for all scales. 
  On the right-hand side we do not see such a change for a down quark in the case of degenerate masses for up and down.}
  \label{Pic:RPud}
\end{figure}

To analyze the effects of parity violation, 
we consider the left-handed  and right-handed  projections of the QPs, 
$\tilde{L}_{AB}$ and $\tilde{R}_{AB}$ respectively. 
They are given by 
\begin{align}
 \tilde{L}_{AB} =& \tilde{A}_{AB} - \tilde{C}_{AB} \left(\propto \gamma^{\mu}(\unit - \gamma^{5})\right)
\nonumber \\
 \tilde{R}_{AB} =& \tilde{A}_{AB} + \tilde{C}_{AB} \left(\propto \gamma^{\mu}(\unit + \gamma^{5})\right).
 \label{Eqn:ProjLR}
\end{align}
It is useful to define their relative ratio $\tilde{r}_{AB}$ as
\begin{align}
 \tilde{r}_{AB}(p^2) &= \frac{\tilde{L}_{AB}(p^2)-\tilde{R}_{AB}(p^2)}{\tilde{L}_{AB}(p^2)+\tilde{R}_{AB}(p^2)}
 = - \frac{\tilde{C}_{AB}(p^2)}{\tilde{A}_{AB}(p^2)}.
 \label{Eqn:RR}
\end{align}
$\tilde{r}$ is positive, if we have dominantly left-handed contributions to the QPs and negative 
for dominantly right-handed contributions. 
$\tilde{A}$ is always positive for pure flavor QPs, which is discussed in \cite{Maas2016prep}. 
Hence the information about the direction of the dominant contribution for QPs is directly related to the sign of $\tilde{C}$.

In the analysis for the axial channel we found a transition scale for the sign change. 
This directly leads to different dominant contributions above and below the transition scale. 
This is depicted in figure \ref{Pic:RPud}. 
On the left hand side we plot the results for physical quark masses of up and down quark. 
At $g_{\text{w}}=10^{-5}$, which is below the threshold strength, $\tilde{C}$ has a sign change for up and down quark. 
Therefore the QP has dominantly right-handed contributions in the UV and dominantly left-handed contributions in the IR. 
Above the threshold strength, the down quark has dominantly right-handed contributions for all scales.

The same is true for the top quark, because $\tilde{C}$ has no sign change for the top quark even at lower breaking strength, 
see figure \ref{Pic:CP} bottom left. The change of dominant contribution for 
the QPs is due to the sign change of the axial channel, which is always present for degenerate masses. 
We conclude that due to the mass splitting of the quarks parity violation in the IR is in 
different directions above the threshold strength. 
The heavier quark has dominantly right-handed contributions and the lighter quark has dominantly left-handed.

Considered from the point of view of the back-coupling, this has an intriguing implication. 
The breaking terms mimic the electroweak interactions, 
while the QCD interaction describes the propagation through the QCD vacuum. 
At tree-level, mass-splittings introduce a (trivial) flip of helicities of the quarks over a sufficiently long distance. 
This helicity flip is suppressed by the back reaction of the QCD vacuum over long distances for the heavier quark flavor, 
if there is a mass splitting and sufficiently strong back-coupling. 
Since the helicity is important for other electroweak interactions, the back-coupling could 
deplete or increase the available amount of quarks in a neutron star, if the same behavior persists in the full case.

\begin{figure}
 \includegraphics[scale=0.9]{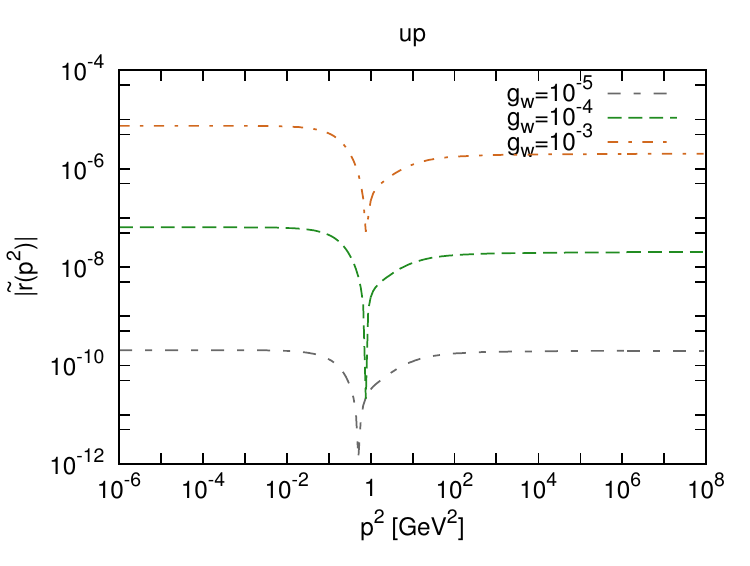}
 \quad
 \includegraphics[scale=0.9]{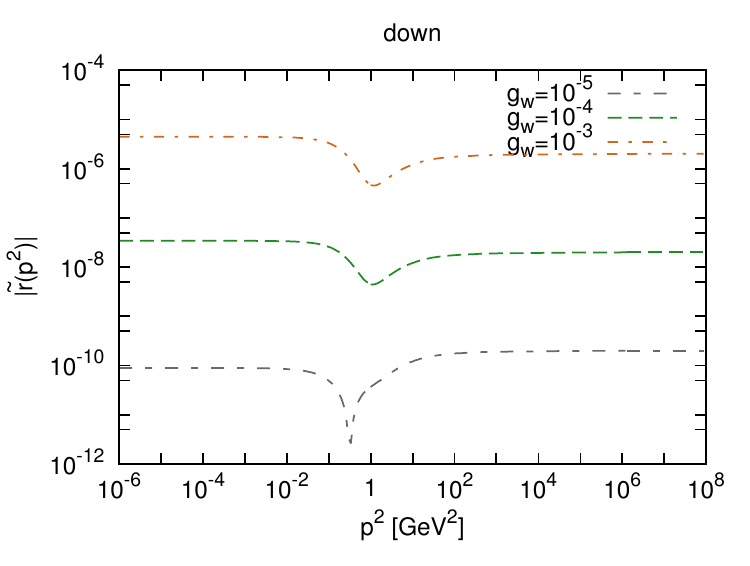}
   \caption{The absolute value of the relative ratio is displayed for the up quark (left-hand side) and down quark (right-hand side). 
   The absolute value in the IR and UV is approximately increased by two orders of magnitude, 
   if the breaking strength is increased by one order of magnitude. Spikes are from the zero crossings in figure \ref{Pic:RPud}.}
  \label{Pic:RPlog}
\end{figure}

However, this non-linear back-coupling effect is not only limited to this helicity flip. 
Consider the absolute value of the relative ratio. It is shown in in figure \ref{Pic:RPlog}. 
Increasing the breaking strength by one order of magnitude increases the absolute value by approximately 
two orders of magnitude for up and down quarks. 
Thus, the effect is non-linearly amplified.

\begin{figure}
 \includegraphics[scale=0.9]{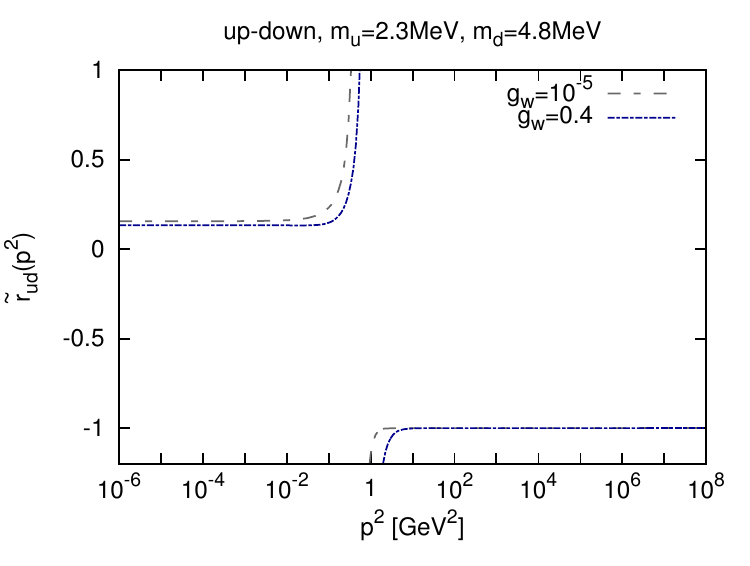}
  \quad
 \includegraphics[scale=0.9]{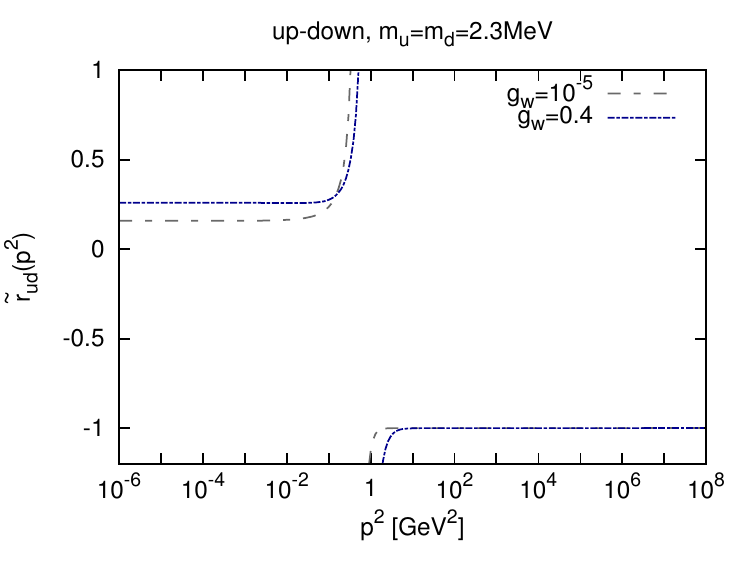}
  \quad
 \includegraphics[scale=0.9]{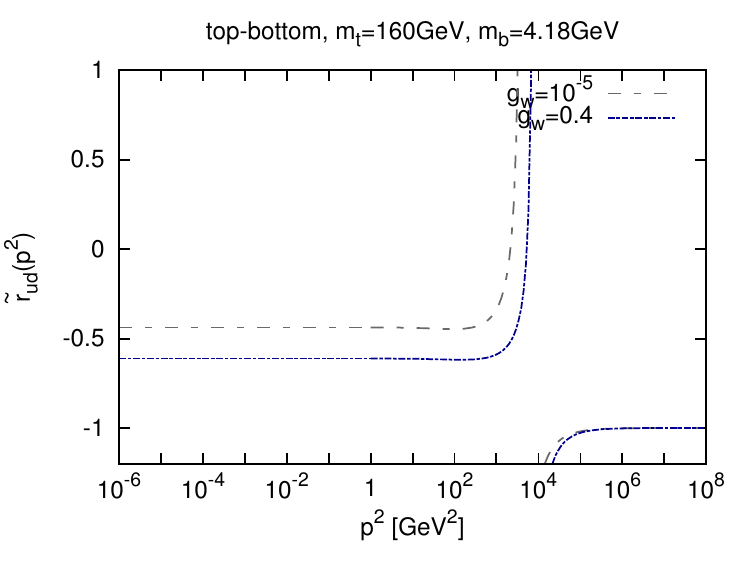}
  \quad
 \includegraphics[scale=0.9]{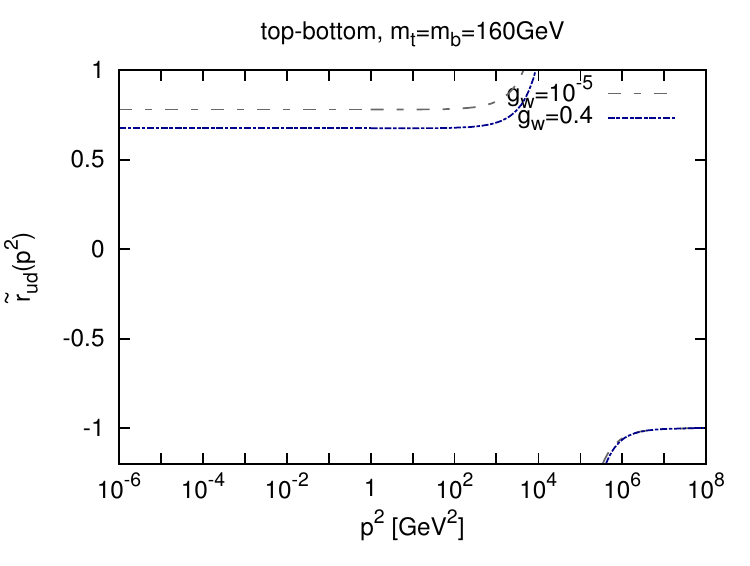}
 \caption{Relative ratios for the QPs of the mixed flavor. 
 QP from up to down quark for the physical quark masses and degenerate masses (top panels) 
 have the dominantly right-handed contributions in the UV and dominantly left-handed contributions in the IR. 
 Increasing the breaking strength does not change the behavior. 
 The QP from top to bottom has dominantly right-handed contributions in the UV and IR, 
 which is due to the bigger value of the mass splitting between top and bottom quarks (bottom panels).}
  \label{Pic:RM}
\end{figure}

Finally, the effects for mixed flavors are shown in figure \ref{Pic:RM}. 
This part of the QP has dominantly right-handed contributions in the UV and dominantly left-handed contributions in the IR. 
The transition appears at about 1 GeV for up and down quarks, which is the same as for the flavor-diagonal ones. 
In this case we find the same behavior for physical quark masses and for degenerate quark masses.

Increasing the strength does not change this pattern, which is different to the flavor-diagonal elements. 
Also, we see only a little quantitative change towards lower values of the breaking for physical up and down quark masses.
In the case of the QPs from top to bottom quark, we see a different behavior (bottom left of figure \ref{Pic:RM}). 
For physical quark masses, this part of the QP has dominantly right-handed contributions in the UV and IR. 
But for the degenerate masses, we still have the same transition from dominantly right-handed to dominantly left-handed, 
which is shown in the bottom right of figure \ref{Pic:RM}. 
Thus for the flavor-mixing part the direction of parity violation depends on the absolute value of the mass 
splitting and not on the the breaking strength, in contrast to the flavor-diagonal case: 
For the small mass splitting of up to down quark the dressing function behaves as for the 
flavor-diagonal elements below the threshold strength. 
For top and bottom the mass splitting is big enough to change the qualitative behavior, 
as it was the case above the threshold strength.

\section{Conclusion}
\label{Conclusion}

We performed a first step on the way to a fully backcoupled and non-perturbative description of the 
electroweak and strong interactions. 
We did this by analyzing the impact of explicit C, P and flavor symmetry violation on the QP. 
We paid particular attention to the influence on parity violation for the the flavor-conserving and flavor-violating propagation.

We found that a threshold scale of breaking exists, 
where the QP changes its dominant contribution from right-handed in the UV to left-handed in the IR. 
This pattern changes for physical quark masses of the mixed flavor independently of the strength of the explicit breaking. 
If the mass splitting was big enough, which was the case for the third generation,
but not for the first generation, then the mixed flavor have dominantly
right-handed contributions in the UV and in the IR.

The pure flavors exhibits a threshold strength in presence of a mass splitting, 
where the behavior changes for the heavier flavor. 
Thus, helicity flips are modified by the backcoupling to QCD at long distances in a qualitative way. 
Remarkably, the threshold strength depends on the absolute value of the mass splitting and is still perturbatively small. 
Thus, non-linear effects substantially amplify the impact of the breaking term.

Thus the dynamical back coupling of the electroweak interactions with the strong interactions need further investigations.
In particular, in systems with a substantial amount of backcoupling, like binary neutron star mergers,
our results could indicate that perturbative calculations may underestimate the importance of the interplay between both sectors.\\

\noindent{\bf Acknowledgements}\\

\noindent We are grateful to Helios Sanchis-Alepuz and Jordi Paris-Lopez for helpful discussions. 
W.\ M.\ has been supported by the FWF doctoral school W1203-N16.

\bibliography{references}

\begin{thebibliography}{16}

\bibitem{Abbott:2016blz}
B.P. Abbott et~al. (Virgo, LIGO Scientific), Phys. Rev. Lett. \textbf{116},
  061102 (2016), \texttt{1602.03837}

\bibitem{Rosswog:2003rv}
S.~Rosswog, M.~Liebendoerfer, Mon. Not. Roy. Astron. Soc. \textbf{342}, 673
  (2003), \texttt{astro-ph/0302301}

\bibitem{Sekiguchi:2011zd}
Y.~Sekiguchi, K.~Kiuchi, K.~Kyutoku, M.~Shibata, Phys. Rev. Lett. \textbf{107},
  051102 (2011), \texttt{1105.2125}

\bibitem{Neilsen:2014hha}
D.~Neilsen, S.L. Liebling, M.~Anderson, L.~Lehner, E.~O'Connor, C.~Palenzuela,
  Phys. Rev. \textbf{D89}, 104029 (2014), \texttt{1403.3680}

\bibitem{Palenzuela:2015dqa}
C.~Palenzuela, S.L. Liebling, D.~Neilsen, L.~Lehner, O.L. Caballero,
  E.~O'Connor, M.~Anderson, Phys. Rev. \textbf{D92}, 044045 (2015),
  \texttt{1505.01607}

\bibitem{Sekiguchi:2015dma}
Y.~Sekiguchi, K.~Kiuchi, K.~Kyutoku, M.~Shibata, Phys. Rev. \textbf{D91},
  064059 (2015), \texttt{1502.06660}

\bibitem{Foucart:2015gaa}
F.~Foucart, R.~Haas, M.D. Duez, E.~O'Connor, C.D. Ott, L.~Roberts, L.E. Kidder,
  J.~Lippuner, H.P. Pfeiffer, M.A. Scheel, Phys. Rev. \textbf{D93}, 044019
  (2016), \texttt{1510.06398}

\bibitem{Caballero:2016lof}
O.L. Caballero, \emph{{Neutrino emission, Equation of State and the role of
  strong gravity}}, in \emph{{11th Latin American Symposium on Nuclear Physics
  and Applications Medellin, Colombia, November 30-December 4, 2015}} (2016),
  \texttt{1603.02755},
  \texttt{https://inspirehep.net/record/1426803/files/arXiv:1603.02755.pdf}

\bibitem{Maas2016prep}
A.~Maas, W.A. Mian, in preparation  (2016)

\bibitem{Alkofer:2000wg}
R.~Alkofer, L.~von Smekal, Phys. Rept. \textbf{353}, 281 (2001),
  \texttt{hep-ph/0007355}

\bibitem{Fischer:2006ub}
C.S. Fischer, J. Phys. \textbf{G32}, R253 (2006), \texttt{hep-ph/0605173}

\bibitem{Alkofer:2008tt}
R.~Alkofer, C.S. Fischer, F.J. Llanes-Estrada, K.~Schwenzer, Annals Phys.
  \textbf{324}, 106 (2009), \texttt{0804.3042}

\bibitem{Roberts:2015lja}
C.D. Roberts, J. Phys. Conf. Ser. \textbf{706}, 022003 (2016),
  \texttt{1509.02925}

\bibitem{Williams:2015cvx}
R.~Williams, C.S. Fischer, W.~Heupel, Phys. Rev. \textbf{D93}, 034026 (2016),
  \texttt{1512.00455}

\bibitem{Maris:1999nt}
P.~Maris, P.C. Tandy, Phys. Rev. \textbf{C60}, 055214 (1999),
  \texttt{nucl-th/9905056}

\bibitem{Goecke:2011pe}
T.~Goecke, C.S. Fischer, R.~Williams, Phys. Lett. \textbf{B704}, 211 (2011),
  \texttt{1107.2588}

\end{thebibliography}

\end{document}